\documentclass[final]{svjour2}
\usepackage{graphicx}
\usepackage{rotating}
\usepackage{amssymb}
\usepackage{mathptmx}
\usepackage[numbers]{natbib}
\makeatletter
\journalname{Journal of Low Temperature Physics}

\bibpunct{}{}{,}{s}{}{,}

\begin{document}

\newcommand{\hdblarrow}{H\makebox[0.9ex][l]{$\downdownarrows$}-}
\title{Comparison of CDMS [100] and [111] oriented germanium detectors}

\author{S.W.~Leman$^1$ \and S.A.~Hertel$^1$ \and P.~Kim$^2$ \and B.~Cabrera$^3$ \and E.~Do~Couto~E~Silva$^2$ \and E.~Figueroa-Feliciano$^1$ \and K.A.~McCarthy$^1$ \and R.~Resch$^2$ \and B.~Sadoulet$^4$ \and K.M.~Sundqvist$^4$ \and on behalf of the Cryogenic Dark Matter Search collaboration}

\institute{1:Massachusetts Institute of Technology, Kavli Institute, \\ Cambridge, MA 02139, USA\\
\email{\texttt{swleman@mit.edu}} \\
2: SLAC National Accelerator Laboratory / KIPAC, \\Menlo Park, CA 94025, USA \\
3: Stanford University, Department of Physics, \\Stanford, CA 94305, USA \\
4: The University of California at Berkeley, Department of Physics, \\Berkeley, CA 94720, USA}

\date{07.20.2011}

\maketitle


\begin{abstract}

The Cryogenic Dark Matter Search (CDMS) utilizes large mass, 3" diameter~$\times$~1" thick target masses as particle detectors. The target is instrumented with both phonon and ionization sensors and comparison of energy in each channel provides event-by-event classification of electron and nuclear recoils. Fiducial volume is determined by the ability to obtain good phonon and ionization signal at a particular location. Due to electronic band structure in germanium, electron mass is described by an anisotropic tensor with heavy mass aligned along the symmetry axis defined by the [111] Miller index (L valley), resulting in large lateral component to the transport. The spatial distribution of electrons varies significantly for detectors which have their longitudinal axis orientations described by either the [100] or [111] Miller indices. Electric fields with large fringing component at high detector radius also affect the spatial distribution of electrons and holes. Both effects are studied in a 3 dimensional Monte Carlo and the impact on fiducial volume is discussed.

\end{abstract}

\section{Introduction}

The present generation of CDMS~\cite{Ahmed2010, Ahmed2011} detectors, deployed at the Soudan Underground Laboratory, are made of high purity germanium crystals with longitudinal axis and major alignment flat orientation defined by [100] and [110] Miller indices respectively~\cite{AshcroftAndMermin}. The detectors are 3~inches in diameter and 1~inch thick with a total mass of about 607~grams. One of the next phases of the CDMS experiment will be deployed in SNOLAB in Sudbury, Canada. In this phase of the experiment we are targeting larger 100~mm diameter, 33~mm thick detectors with interleaved ionization channels~\cite{Brink2006, Oed1988}. Due to details in the electron charge transport we are considering the use of [111] oriented detectors with [2$\overline{1}\overline{1}$] flats. It is the difference in the detector response for these two crystal orientations and the effect on fiducial volume which is the focus of this paper.
		
\section{Charge Transport}

\begin{figure}[ht]
\begin{center}
\includegraphics[width=8cm]{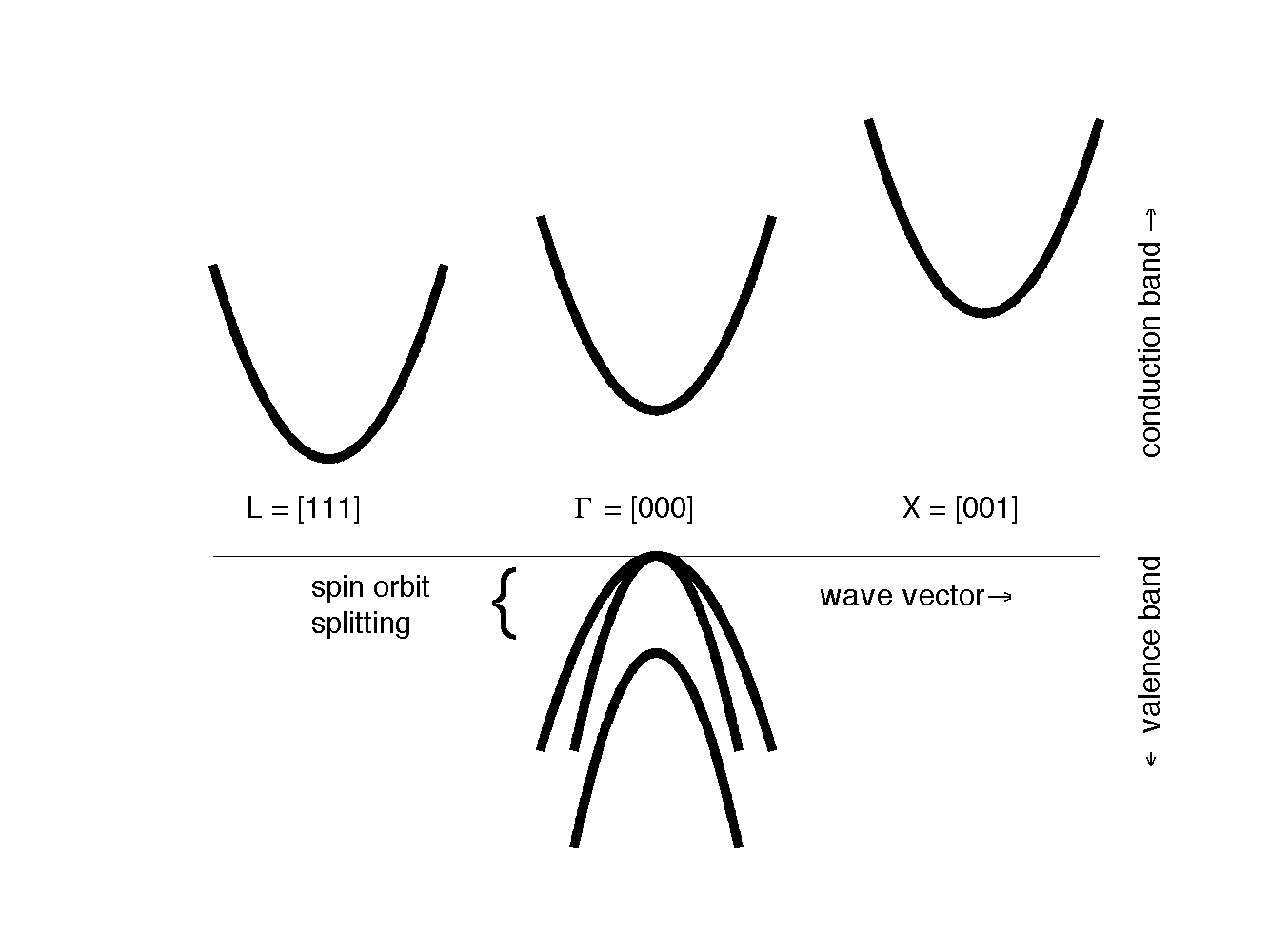}
\end{center}
\caption[] { \label{fig:bandGe} Germanium band structure showing the hole ground state, $\Gamma$ band and electron ground state L bands.}
\end{figure}

Electron transport is described by a mass tensor, leading to electron transport which contains components oblique to the applied field~\cite{Sasaki1958, Jacoboni1983, Leman2011_5} and is necessary to explain and interpret signals in the primary and guard-ring ionization channels which function as a fiducial volume cut. Germanium has an anisotropic band structure described schematically in Figure~\ref{fig:bandGe}. At low field, and low temperature, germanium's energy band structure is such that the hole ground state is situated in the $\Gamma$ band's [000] location and the electron ground state is in the L-band [111] location~\cite{Jacoboni1983}. Hole propagation dynamics are relatively simple due to propagation in the $\Gamma$ band and the isotropic energy-momentum dispersion relationship $\epsilon_{hole}(\mathbf{k}) = \hbar^2 k^2 / 2m$. Electron propagation dynamics are significantly more complicated due to the band structure and anisotropic energy-momentum relationship. At low fields and low temperatures, electrons are unable to reach sufficient energy to propagate in the $\Gamma$ or X-bands, and are not considered in the Monte Carlo. The electron energy-momentum dispersion relationship is anisotropic and given by $\epsilon_{electron}(\mathbf{k}) = \hbar^2 /2 (k_{\parallel}^2 / m_{\parallel} + k_{\perp}^2 / m_{\perp})$, where the longitudinal and transverse mass ratio $m_\parallel/m_\perp \sim$19.5. Modeling of the transport processes is complicated by anisotropy and is handled in the Monte Carlo by a Herring-Vogt~\cite{Herring1956, Cabrera} transformation.
  
For a [100] oriented crystal with field in the z = [100] direction, the charges will propagate in four different lobes, each 33 degrees from the z axis. For a [111] oriented crystal with field in the z = [111] direction, one lobe will propagate along the z axis while the other three propagate 18 degrees from the z axis. The tighter spatial distribution of electrons in the [111] oriented crystal would seem to make this detector type more favorable with less electrons propagating into the side walls and producing a non-bulk detector response, however this benefit is not borne out for two reasons. First, at high radius there are large fringing components to the electric fields which tend to also drive charge carriers (regardless of crystal orientation) into the side walls. Second, in the [111] oriented crystal, the [111] valley has light mass in the orthogonal directions and the fringing fields result in lateral transport towards the detector side walls. A Monte Carlo of charge transport with an accurate electric field model is required to understand the tradeoffs between these effects in the different crystal orientations.

\begin{figure}[ht]
\begin{center}
\hspace*{-0cm}\includegraphics[width=11cm]{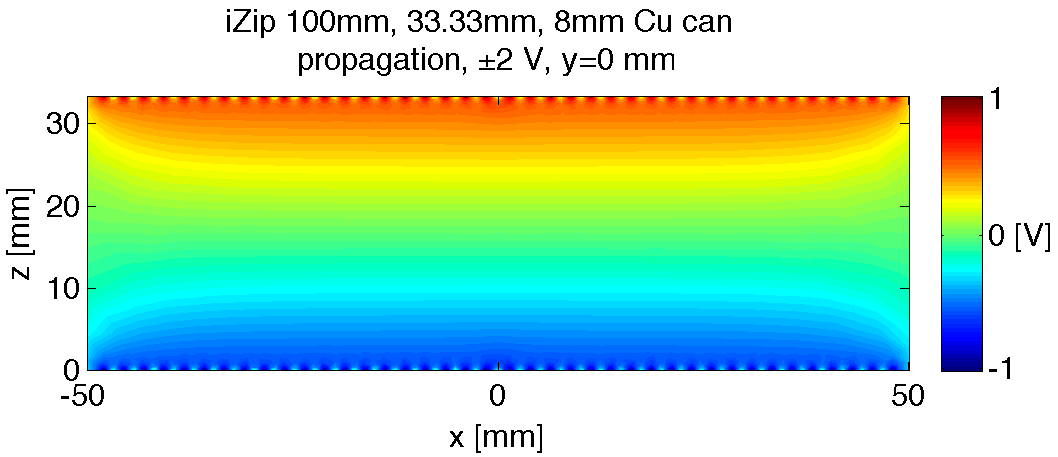}
\end{center}
\caption[] { \label{fig:EPot} Electric potential for a 100~mm diameter, 33.33~mm thick iZIP detector. The detector is biased at $\pm$2~V but the color scale in the plot is $\pm$1~V to emphasize the fringing fields at high radius.}
\end{figure}

A map of the electric potential is shown in Figure~\ref{fig:EPot} for a 100~mm diameter, 33.33~mm thick iZIP detector, the largest detector size that we are targeting for the CDMS-SNOLAB phase of the experiment~\cite{BrinkLTD14}. It is noticed that there is a significant horizontal component to the gradient (electric field) which results in lateral transport of charges into the detector's sidewall. 

Monte Carlo were run with combinations of detector dimension, crystal orientation and germanium-encasing copper structure distance as shown in Table\ref{tab:fidVol}. The Monte Carlo results were then processed in the CDMS Analysis Package in a fashion identical to data from the experiment. A part of that analysis involves imposing fiducial volume cuts and determining the ionization collection efficiency. In the WIMP search mode, the ionization to phonon energy ratio is used to discriminate between gamma and neutron events where neutron recoils produce about 1/3 the ionization to phonon ratio as gammas of the same energy~\cite{Lindhard1963}. For gamma recoils the charge collection efficiency is $<$1 in some regions of poor detector response near the detector surfaces. It is critical that the fiducial volume cuts eliminate regions of reduced charge collection to ensure that background gamma events do not leak into the nuclear recoil signal region. This fiducial volume is defined such that there is no signal in the outer charge guard rings (or more precisely, the outer channels' signals are consistent with noise) in either the top or bottom charge channels. Additionally, there is a requirement that the signals in the top and bottom charge channels are within 4.5-5.6\% of each other (depending on the detector geometry), indicating an interaction in the detector bulk.

\begin{figure}[ht]
\begin{center}
\hspace*{-1cm}\includegraphics[width=13cm]{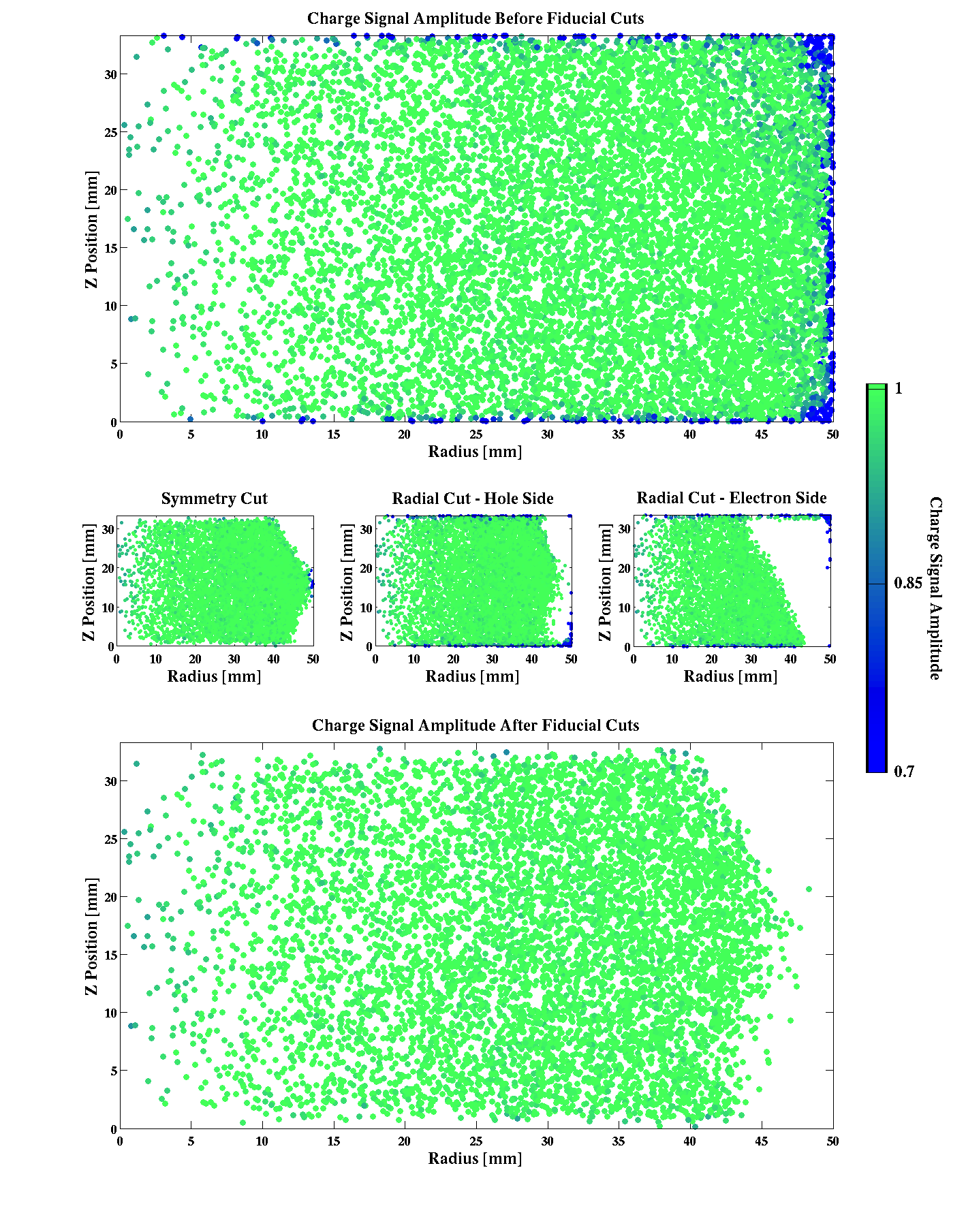}
\end{center}
\caption[] { \label{fig:fidVol} Charge signal amplitude for 10,000 events randomly distributed in a 100~mm diameter, 33.33~mm thick germanium detector with [100] orientation. The point represents the location of energy deposition and the color indicates the charge signal amplitude. The top plot is before any cuts, the middle three plots for three independent cuts and the bottom plot after the cuts are applied. For an event to be removed by the radial cut, it must fail both the electron and hole radial cuts.}
\end{figure}

The interaction location for 10,000 gamma events is shown in Figure~\ref{fig:fidVol} where the color of each point represents the charge signal amplitude. Events at large radius or near the top or bottom detector faces result in a reduced charge signal amplitude as discussed earlier. The symmetry cut itself is very effective at removing low-charge events near the top and bottom channels, by design. The symmetry cut is also effective at removing low-charge events at large radius since usually only one carrier type will trap on a sidewall and result in an asymmetric signal. The exception is however near mid elevation where both carrier types trap on the sidewall with similar losses and these low-charge events leak past the symmetry cut. These large radius events do fortunately result in a strong signal in either / both of the electron-side or hole-side outer guard channels and are rejected. The hole-side radial cut shows a fiducial volume region whose shape is largely determined by fringing fields, whereas the electron-side radial cut shows a region whose shape is largely determined by the electron transport oblique to the local field.

The fiducial volumes are described in Table~\ref{tab:fidVol}, where a germanium density of 5.35~g~cm$^{-3}$ was used to compute the mass and the detectors were assumed to be cylindrical (no mass was removed for alignment flats). There are two striking and counter-intuitive findings of these Monte Carlo runs. The first is that the [111] oriented crystals are no better than the [100] variety. While it is true that the electron transport trajectories are different, the outer radius fiducial cut is selected by hole transport trajectories. The charge symmetry cut is weakly affected by the electron transport trajectories but only weakly, leading to comparable fiducial volumes in the two crystal orientations. The second unexpected result is that a decrease in the radius of the copper can surrounding the germanium leads to an increased fiducial volume. In this geometry, events at mid elevation are directed into the inner ionization channel by the large fringing fields.

\begin{table}[ht]
\caption{Detector geometries and crystal orientation and germanium-copper ground can radial spacing tested in Monte Carlo along with the resulting fiducial volumes after radial and symmetry cuts are applied. The crystal orientation refers to the crystal symmetry axis that points along the detector's longitudinal axis.}
\begin{center}
\begin{tabular}{|c|c|c|c|c|c|}
\hline
diameter $\times$ & orientation & Cu can          & Ge-Cu    & fiducial       & fiducial   \\ 
 thickness               &                     & diameter        & space  & volume  & volume    \\ 
 ~[mm] &                          &  [mm] &  [mm] &  [g]  &  [\%]   \\ 
\hline
76.2 $\times$ 25.4 & 100 & 80.2 & 2 &  400  & 64 \\
76.2 $\times$ 25.4 & 100 & 84.2 & 4 & 380  & 61\\
76.2 $\times$ 25.4 & 100 & 92.2 & 8 & 360  & 58\\
\hline
100 $\times$ 25.4   & 100 & 104 & 2 & 720  & 67\\
100 $\times$ 25.4   & 100 & 108 & 4 & 700  & 66\\
100 $\times$ 25.4   & 100 & 116 & 8 & 700  & 65\\
\hline
100 $\times$ 33.33 & 100 & 104 & 2 & 980  & 70\\
100 $\times$ 33.33 & 100 & 108 & 4 & 950  & 68\\
100 $\times$ 33.33 & 100 & 116 & 8 & 920  & 66\\
\hline
76.2 $\times$ 25.4 & 111 & 80.2 & 2 & 380 & 61\\
76.2 $\times$ 25.4 & 111 & 84.2 & 4 & 380  & 61\\
76.2 $\times$ 25.4 & 111 & 92.2 & 8 & 370  & 59\\
\hline
100 $\times$ 25.4   & 111 & 104 & 2 & 700  & 66\\
100 $\times$ 25.4   & 111 & 108 & 4 & 700  & 66\\
100 $\times$ 25.4   & 111 & 116 & 8 & 700  & 65\\
\hline
100 $\times$ 33.33 & 111 & 104 & 2 & 950  & 68\\
100 $\times$ 33.33 & 111 & 108 & 4 & 950  & 68\\
100 $\times$ 33.33 & 111 & 116 & 8 & 940  & 67\\
\hline
\end{tabular}
\end{center}
\label{tab:fidVol}
\end{table}

\section{Conclusions}

A detailed charge MC of CDMS detectors in various geometry and crystal orientations has shown that [100] oriented crystals have fiducial volume comparable to those of the more expensive [111] variety. Furthermore, the fiducial volume increases for a decreased germanium-copper ground can spacing in the radial direction. 

\begin{acknowledgements}

This work is supported by the United States National Science Foundation under Grant No. PHY-0847342.

\end{acknowledgements}

\pagebreak

\end{document}